# Data science and AI in FinTech: An overview

**Longbing Cao** · **Qiang Yang** · **Philip S. Yu**

**Abstract** Financial technology (FinTech) has been playing an increasingly critical role in driving modern economies, society, technology, and many other areas. Smart FinTech is the new-generation FinTech, largely inspired and empowered by data science and new-generation AI and (DSAI) techniques. Smart FinTech synthesizes broad DSAI and transforms finance and economies to drive intelligent, automated, whole-of-business and personalized economic and financial businesses, services and systems. The research on data science and AI in FinTech involves many latest progress made in smart FinTech for BankingTech, TradeTech, LendTech, InsurTech, WealthTech, PayTech, RiskTech, cryptocurrencies, and blockchain, and the DSAI techniques including complex system methods, quantitative methods, intelligent interactions, recognition and responses, data analytics, deep learning, federated learning, privacy-preserving processing, augmentation, optimization, and system intelligence enhancement. Here, we present a highly dense research overview of smart financial businesses and their challenges, the smart FinTech ecosystem, the DSAI techniques to enable smart FinTech, and some research directions of smart FinTech futures to the DSAI communities.

**Keywords** Financial technology · FinTech · Financial service · Smart FinTech · Finance · Economics · Blockchain · Data science · AI · Intelligent systems · Machine learning · Deep learning · Federated learning · Privacy-preserving · Modeling · Mathematics · Statistics · Ethics

## 1 Introduction

In recent years, finance has become increasingly interactive with new-generation data science and artificial intelligence (DSAI) techniques [2, 15, 7, 27, 48, 49]. In particular, FinTech (or Fintech) [13, 44] is at the epicentre of synthesizing, innovating and transforming financial services, economy, technology, media, communication, and society broadly driven by DSAI techniques [8]. Here, *DSAI* broadly refers to (1) *classic AI areas* including logic, planning, knowledge representation, modeling, autonomous systems, multiagent systems, complexity science, expert system (ES), decision support system (DSS), optimization, simulation, pattern recognition, image processing, and natural language processing (NLP); (2) *advanced DSAI areas* including intelligent interactions and conversations, intelligent identification and authentication, privacy-preserving processing [19], advanced representation learning, advanced analytics and learning, knowledge discovery, computational intelligence, event and behavior analysis, social media/network analysis; and (3) more recent advances including deep learning [22], automated interactions, learning and responses, transfer learning [47], federated learning [57, 56], human-centered computing, and brain and cognitive computing [42]. Other fundamental areas such as statistical modeling and mathematical modeling also play a critical role in enabling FinTech. In contrast, our reference to *finance* includes areas such as capital market, trading, banking, insurance, leading/loan, investment, wealth management, risk management, marketing, compliance and regulation, payment, contract, auditing, accounting, and digital currencies and their supporting infrastructure (including blockchain [45]), operations, services, management, security, and ethics. Economics and finance (EcoFin) are increasingly synergized with FinTech and DSAI.

DSAI is the keystone enabler of the new generation of EcoFin and FinTech [15, 28, 35, 39]. The new-generation DSAI is reshaping or even redefining the concepts, objectives, content and tasks of EcoFin and FinTech. DSAI essentially and comprehensively transforms the way that modern economic and financial (economic-financial) businesses operate, transact, interact and collaborate with their participants (incl. consumers, markets, and regulators) and environments. DSAI nurtures new economic-financial mechanisms, models, products, services, and many tangible and intangible opportunities. As a result, DSAI not only strengthens the efficiency, cost-effectiveness, customer experience, risk mitigation, regulation, and security of existing economic-financial systems, it also innovates unprecedented and more intelligent, efficient, convenient, personalized, explainable, secure and proactive economic-financial products and services. The synthetic product of DSAI and EcoFin forms the new area of *smart FinTech*, as shown in Fig. 2 in [8], which presents a four-dimensional, systematic and interactive landscape of the synthesis between DSAI and EcoFin in forming smart FinTech. The landscape connects the main EcoFin businesses to the EcoFin data and repositories, the broad-based DSAI techniques, and the EcoFin business objectives. Accordingly, the family of FinTech has expanded from BankingTech, TradeTech, LendTech, InsurTech, WealthTech and PayTech to RiskTech, etc.

In the rest of this paper, we discuss and summarize the main business domains and their challenges, the ecosystems of smart FinTech which substantially expand the above FinTech family, and the DSAI techniques driving smart EcoFin businesses and FinTech. A brief introduction to this selected topic on data science and AI in FinTech is then given, followed by discussion on future directions.

## 2 FinTech businesses and challenges

Below, we briefly summarize the main businesses and challenges in smart FinTech[1]. Typical applications of smart FinTech include Internet banking, mobile payments, online shopping, peer-to-peer leading, online crowdfunding projects, cryptocurrency, cross-market portfolio management, and global supply chain management.

**FinTech business areas.** Building on DSAI techniques, these areas involve almost all aspects of an economic-financial system and its environment and broadly all EcoFin businesses [8, 9, 2, 39]. Here, we highlight the following major business areas in smart FinTech:

- economic-financial innovations: e.g., new mechanisms and products;
- economic-financial markets: including products and services;
- economic-financial participants: including individual and retail investors, institutions and regulators;
- economic-financial behaviors: e.g., investor activities and company announcements;
- economic-financial events: e.g., company mergers and financial crises;
- economic-financial services: e.g., banking, insurance, lending, financing, and crowdfunding services;
- economic-financial mechanisms: e.g., market mechanisms, business models, and derivative pricing;
- economic-financial systems: e.g., a company's finance;
- economic-financial infrastructure: including fundamental support systems and blockchain;
- economic-financial pricing: including the valuation of underlying and derivative assets in capital markets;
- economic-financial trading: including trading, investment, and execution;
- economic-financial payment: e.g., online, mobile and contactless payment;
- economic-financial valuation: e.g., the estimation of property values, credit and intangible assets;
- economic-financial marketing: including campaign, and customer care;
- economic-financial relationship management: e.g., stakeholder relations, and business partnerships;
- economic-financial resource management: including the management of human, material, informational and intangible assets;
- economic-financial operations: e.g., the processes and services for supporting financial innovation, design and production;
- economic-financial compliance and regulation: e.g., the enforcement of operational orders and business rules by the authorities;
- economic-financial crisis, risk and security: e.g., investment risk, systemic risk, and cybersecurity; and
- economic-financial ethics: including social and ethical issues and privacy.

**FinTech challenges.** Major challenges associated with the above FinTech business areas as well as their associated EcoFin businesses, problems, data and objectives can be categorized into:

- innovation challenges: e.g., DSAI techniques for inventing novel, efficient, intelligent and sustainable mechanisms, products, services and platforms;
- business complexities: such as DSAI techniques for representing, learning and managing the intricate

[1] Interested readers can refer to a comprehensive review on AI in finance in [8,9].

working mechanisms, structures, interactions, relations, hierarchy, scale, dynamics, anomaly, uncertainty, emergence and exceptions associated with a market, a product or a participant;
– organizational and operational complexities: such as DSAI techniques for understanding and managing the diversity and personalization of individuals and departmental teams, departmental and institutional coherence and consensus, and inconsistent and volatile efficiency and performance;
– human and social complexities: such as DSAI techniques for modeling and managing the diversity and inconsistency of a participant's cognitive, emotional and technical capabilities and performance, and for enabling effective communications, cooperation and collaboration within a department and between stakeholders;
– environmental complexities: such as DSAI techniques for modeling and managing the interactions with contextual and environmental factors and systems and their influence on a target business system and problem;
– regional and global challenges: such as understanding and managing the relations between an economy entity and its financial systems with the related regional and global counterparts and stakeholders and their influence on target problems;
– data complexities: such as extracting, representing, analyzing and managing data quality issues, misinformation and complicated data characteristics, e.g., uncertainty and extreme dimensionality, sparsity and skewness;
– dynamic complexities: such as modeling, predicting and managing evolving but nonstationary behaviors, events and activities of individual and block markets, products, services and participants; and
– integrative complexities: e.g., systematically modeling and managing the various aspects of the above complexities that are often tightly and loosely coupled with each other in an underlying economic-financial system.

## 3 The smart FinTech ecosystem

FinTech is evolving to be a large family, comprising various areas including BankingTech, TradeTech, LendTech, InsurTech, WealthTech, PayTech, and RiskTech. The smart FinTech ecosystem is a multi-dimensional synergy between EcoFin business objectives, business areas, data and resources, and enabling techniques, as shown in Fig. 2 in [8] which presents a four-dimensional landscape of the smart FinTech and the synthesis between these aspects. In general, the ecosystems of smart FinTech can be viewed in terms of their (1) comprehensive FinTech-driven businesses and areas, and (2) processes, functions and activities. On one hand, FinTech-driven businesses and areas are overwhelming and evolving. Fig. 1 overviews the smart FinTech ecosystems and categorizes them as follows: smart banking, smart insurance, smart lending, smart trading, smart assets and wealth management, smart payments, smart credit and risk management, smart Internet finance, smart marketing and services, smart currencies and blockchain, smart compliance and regulation, smart ethics and security, and smart innovations. Each of these primary areas is further defined by their respective businesses and supporting (DSAI) techniques [9, 27, 39, 44]. Below, we briefly illustrate how DSAI techniques can assist in these smart FinTech areas.

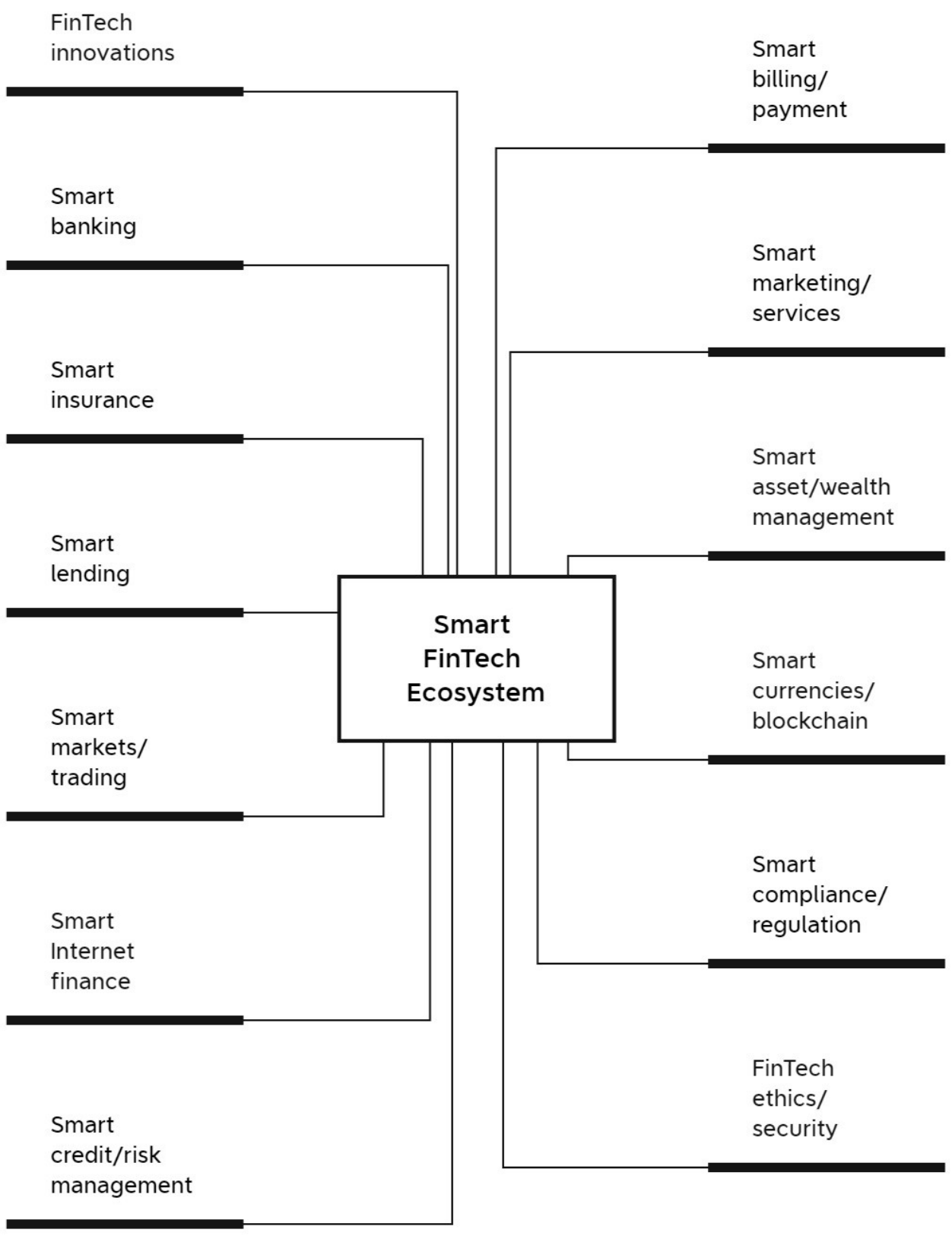

**Fig. 1** The Smart FinTech ecosystem.

**Smart banking** offers active, personalized, automated, trusted, robust, secure, risk-averse and fragility-averse banking businesses and services, including retail banking, open banking, mobile banking, and business banking. The DSAI research for enabling smart banking includes DSAI techniques for detecting, analyzing and managing a bank's network risk and risk contagion; modeling, detecting and managing banking risk and fraud; conducting credit analysis and improving pricing; optimizing, validating and risk-managing credit loans; creating, mining, securing, risk-managing and optimizing digital currencies including cryptocurrencies; creating, securing, evaluating, risk-managing and optimizing mobile banking businesses and services; personalizing, automating, validating, securing, risk-managing and optimizing Internet/online banking; enabling, personalizing, automating, securing, risk-managing and optimizing open banking; evaluating, securing and risk-managing shadow banking; and developing smarter and more automated, personalized, mobile and engaging banking services; etc.

**Smart insurance** enables insurance products, systems and services to ensure safe/secure, cost-effective, proactive, tailored, trustful, resilient, secure and risk-averse health, car, home/content/building, travel and other businesses. The DSAI research for enabling smart insurance includes those for tailoring individual, business and commercial insurance products and services; enabling early, active and evolving insurance fraud detection; making active, personalized and time-varying recommendations of insurance products and services; evaluating, analyzing, detecting, managing and optimizing insurance risk and compliance; evaluating, automating, detecting and optimizing insurance security; creating novel insurance models, products and services; evaluating, automating and improving insurance business operations; and integrating, recommending and optimizing multi-policy, multi-product and cross-selling insurance products and services; etc.

**Smart lending** supports lending, loan and mortgage businesses and services that are risk-averse, personalized, context-oriented, predictive, efficient, robust and secure for individuals, enterprises or projects. DSAI research directions for smart leading cover businesses and tasks such as for addressing relevant aspects and problems through blockchain; automating crowdfunding, e.g., campaign design and strategy optimization; creating and optimizing distributed ledger technologies; enabling fundraising such as by creating personalized recommendations, efficient fundraising models, and intelligent services; analyzing, detecting, managing and mitigating lending risk and security; enabling smart, efficient and low-risk peer-to-peer lending; and creating, validating and optimizing smart contracts; etc.

**Smart markets and trading** involves very broad EcoFin businesses and widely-explored DSAI techniques for intelligent capital, commodity, currency, energy, commerce, property and other markets, and supplies predictive, active, dynamic, risk-averse, anti-fragile, and high-utility trading strategies, support and services. The DSAI-driven smart markets and trading can support many tasks, e.g., for designing, automating and optimizing novel, secure, smart and personalized financial mechanisms, models, products, and services for equity and derivative markets; making and optimizing algorithmic trading, arbitrage trading, high frequency (cross-market) trading, and institutional trading across equity, commodity and currency markets; modeling, automating and optimizing digital asset pricing and credit scoring; enabling smart e-commerce for personalized retail businesses; enabling smart, efficient, personalized and secure Internet finance; conducting portfolio analytics and enabling smart portfolio management; enabling predictive trading and strategic trading; designing, automating and optimizing trading strategies, algorithms and platforms; enabling and optimizing socially responsible investment; modeling, evaluating and optimizing trading complementarity and substitutionary; and characterizing and improving trading incentives and campaign; etc.

**Smart Internet finance** has evolved from Internet banking to third-party payment, peer to peer lending (P2P lending), crowdfunding, and digital currencies, which also involve their supporting designs, operations, services, evaluation, compliance, and security (trust). More broad Internet financial services include Internet platforms, data and technology-driven financing, fund, lending, insurance, payment and markets associated with individuals, corporate organizations, sponsors, and regulators. Typical FinTech for smart Internet finance includes inventing new Internet technologies and services to enable the above products and services; enabling efficient, trustful, secure and convenient Internet banking services, authentication, active compliance checks, outlier detection, and connections to other banking businesses such as mobile banking; enabling equity, product and reward-based crowdfunding mechanisms, projects and their developments and risk management, e.g., default prediction, evaluating the success and risk of crowdfunding projects, automatic screening of illegal fundraising, evaluating and managing herding effect, social networking and mutual influences, and advising government regulation on investors and corporate; optimizing P2P lending bidding mechanisms and loan evaluation, predicting and managing P2P lending herding effect and information asymmetry, and evaluating their influence and systemic risk on investment and the economy; optimizing pricing, price volatility and fluctuation, payment accuracy, risk on investment, regulation, operation security and efficiency of digital currencies. Other areas where DSAI may contribute include enabling, securing, and regulating so-called big data finance, information-driven financial institutions, and smart systems and services for Internet finance.

**Smart billing and payment** offers efficient, secure, risk-averse, fast and convenient systems and services for online, mobile, WiFi, contactless (including for credit cards and by QR-codes) and IoT-oriented billing and payment services. Smart billing and payment becomes increasing important in the digital society and economy. DSAI-enabled smart payment research include areas such as authenticating, automating, securing and risk-managing billing validity and contactless payment; supporting efficient and secure IoT device-based contactless payments; enabling, securing, risk-managing and optimizing e-payments; automating, securing, risk-managing and optimizing Internet and online payments; automating, securing, risk-managing and optimizing mobile payments; modeling, detecting, analyzing and mitigating payment risk; and validating, detecting, analyzing and mitigating billing and payment risk and security; etc.

**Smart credit/risk management** offers efficient, personalized, active, evolving, secure and sustainable credit and risk valuation, products, services and management for business and customers. Typical DSAI-enabled smart credit and risk management tasks include automating, evaluating and optimizing anti-money laundering; monitoring, detecting, categorizing, factorizing, predicting and intervening client financial security, retail investor risk, financial systematic risk, financial institutional risk, financial network risk, and cross-market risk; detecting, quantifying and predicting financial risk factors and areas; quantifying, analyzing, detecting, predicting and mitigating financial crisis and crisis contagion; categorizing, monitoring, detecting, analyzing, predicting, evaluating and managing financial events, market movement, change, exception and emergence and their consequences and impact; etc.

**Smart asset and wealth management** offers businesses and services safe, secure, personalized, anti-fragile, and automated management of money, credit, properties, securities and intangible assets. DSAI-supported smart asset and wealth management involves many emergent research areas, including DSAI for making, evaluating and optimizing data monetization; valuating, analyzing, evaluating, optimizing and managing digital assets, Internet wealth, public welfare, institutional welfare, and superannuation; enabling, automating, risk-managing and optimizing digital financial advising; designing and optimizing novel, personalized, secure and healthy financing mechanisms; enabling, automating, evaluating and optimizing roboadvising; and supporting secure, personalized, low-risk and sustainable venture capital management; etc.

**Smart currencies and blockchain** enables efficient, secure, risk-tolerant, automated or semi-automated, dynamic, and high-performing blockchain infrastructures, computing and services for encrypted digital currencies. DSAI techniques can play a critical role in enabling smart blockchains. Examples are predicting the price movement of cryptocurrencies; constructing risk-averse trading strategies and portfolios of bitcoins and other cryptocurrencies; detecting risk and assuring smart contracts; enabling the efficient mining of bitcoins in distributed systems; supporting proactive and systemic blockchain governance and regulation; and detecting risk, intrusion and fragility in blockchain systems, behaviors and activities; etc.

**Smart marketing and services** offers cost-effective, relevant, proactive, personalized, positive, sequential and context-oriented marketing, advertising and recommendation activities and services. DSAI-enabled smart marketing involves many opportunities, such as conducting consumer sentiment and public emotion analysis; understanding consumer/client opinion and intention; quantifying, characterizing, evaluating, predicting and improving consumer confidence and a product/service's market reputation and trust; evaluating and optimizing customer relationship management; validating, enhancing, synergizing and optimizing econometrics; evaluating, risk-managing, personalizing and optimizing stakeholder relationships; making, evaluating and optimizing financial recommendations; enabling, automating, evaluating, securing, risk-managing and optimizing supply chain finance; and supporting user privacy-preserving and advertiser information-protected federated advertising; etc.

**Smart compliance and regulation** supports automated or human-machine-cooperative, risk-sensitive, proactive, systematic, dynamic and evidence-based regulation and compliance operations, governance and risk intervention and control. The opportunities associated with DSAI-empowered smart compliance and regulation are comprehensive and significant for any FinTech business and technologies, such as enhancing corporate finance transparency; evaluating and optimizing corporate governance and regulation; making and optimizing cross-market regulation; creating and optimizing digital currency regulation; detecting, analyzing, risk-managing and managing financial crime; verifying, automating, securing and optimizing financial digital authentication; enabling, validating, verifying, detecting, risk-managing and optimizing financial digital identification; quantifying, validating, monitoring, detecting, analyzing and mitigating financial fragility, crisis and stability; recognizing, detecting, analyzing and mitigating financial fraud; enabling smarter and more efficient and personalized financial market regulation, design and policy implication; automating, eval-uating and optimizing financial operations; quantifying, evaluating and managing information asymmetry and transparency; enabling, automating and optimizing international regulation; validating and improving market legitimacy; quantifying, verifying, detecting, analyzing and improving market social justice; and quantifying, evaluating, automating and improving marketplace trust and coordination; etc.

**FinTech ethics and security** involves whole-of-businesses, whole-of-time, whole-of-stakeholders, whole-of-physical-cyber-spaces, privacy-preserving, proactive, predictive, systematic, and dynamic management of FinTech ethics and security. DSAI techniques can help with extracting evidence and profiling ethical and security violations; modeling, detecting, evaluating and managing financial system security, financial network security, and financial instrument security; characterizing, quantifying, analyzing, evaluating, predicting and managing financial system vulnerability; and providing distributed, federated, privacy-preserving financial services, etc.

**FinTech innovations** drive smart FinTech developments and applications. Every area of smart FinTech needs continuous innovation and research, where DSAI technology is essential. Examples of DSAI-driven FinTech innovations are data and learning-enabled systems and services for automated pricing, credit scoring, loan valuation, trading strategy generation, customer chatbots, financial planning, security alerting, and compliance mitigation; proactive and personalized recommendations of crowdfunding projects, cross-product loan, insurance and investment portfolios and pricing; tailored risk-mitigated systems and services for asset and wealth portfolio prediction, optimization and risk management for high-value customers; detecting and intervening unethical and unsecure trading, lending, credit and loan valuation, payment, marketing, competition, and regulation; offering whole-of-business, privacy-preserving and federated FinTech and EcoFin businesses and services to large-scale, distributed and connected communities or societies, etc.

On the other hand, all of the above smart FinTech businesses and areas involve some common fundamental processes, functions and activities. Fig. 2 illustrates the main processes and their key functions and activities: design, produce, operate, promote, optimize and safeguard.

- Design: plan and devise EcoFin mechanisms, markets, products, participants, rules and governance, and innovations, etc.
- Produce: convert and implement the designs into production with goods, services, systems and applications and functions and activities including pricing and valuation, trading and exchange, supply and demand, financial services (e.g., billing and payment), etc.
- Operate: enable and support the whole-of-business execution by providing technology (including infrastructure) and managing resources (money, asset, personnel and management, data, facilities, etc.), communications (internal and external), stakeholders, processes, and operations, etc.
- Promote: market and advocate businesses and opportunities, including supporting advertising, competition, recommendations, new developments (e.g., clients, products, channels, etc.) and managing stakeholder relationships, etc.
- Optimize: evaluate, improve and expand the quality and performance (including efficacy and efficiency) of EcoFin businesses and operations and make optimal planning, strategies, reinforcement, recommendations and innovations, etc.
- Safeguard: validate, secure and assure the compliance, regulation, privacy, security, trust and ethics etc. of EcoFin businesses and FinTech.

## 4 DSAI techniques enabling smart FinTech

The DSAI techniques required to achieve the aforementioned smart FinTech objectives and EcoFin businesses are very broad and diversified [6, 11, 15, 16, 20–22, 24, 26, 27, 31, 33, 40, 41, 46]. There are different ways to categorize the techniques enabling smart FinTech. For example, Fig. 2 [9] categorizes the techniques into (a) mathematical and statistical modeling, (b) complex system methods, (c) classic analysis and learning methods, (d) computational intelligence methods, (e) modern DSAI methods, (f) deep financial modeling methods, and (g) hybrid methods. Table 1 details the popular techniques and their applications for smart EcoFin businesses and FinTech.

Here, we expand the above scope and briefly discuss various intelligent services required for smart EcoFin and FinTech. The typical categories and applications of various DSAI techniques for smart FinTech include:

- complex system methods: techniques to characterize and design complex EcoFin systems;
- quantitative methods: techniques to quantify EcoFin designs and businesses;
- intelligent interactions and responses: to enable intelligent human-machine interactions and effective conversations with stakeholders and autonomous or personalized responses;

- analytics and learning techniques: to analyze, evaluate, optimize and recommend smart FinTech businesses, services, and data;
- optimization and augmentation techniques: to evaluate, enhance and expand the quality, performance and opportunities of smart businesses, data and FinTech;
- deep financial modeling: to involve deep, reinforced, transfer and Bayesian learning techniques and other advances in the representation, analysis, prediction and management of financial businesses and services;
- system intelligence enhancement: to enhance autonomous and intelligent interfaces, technologies, infrastructures and systems for smarter EcoFin businesses and FinTech; and
- whole-of-business privacy-preserving federated FinTech: to provide unified but distributed secure financial services by connecting the entire aspects of humans, systems, businesses, IoT devices, and data.

All of these techniques may be applied in each of the aforementioned six processes and functions of smart FinTech. Each technique may play a distinct role in addressing the respective EcoFin business challenges and data complexities, as illustrated below.

**Complex system methods.** The design of EcoFin businesses and smart FinTech often relies on and involves theories of complex systems and social sciences, including brain and cognitive computing, complexity science methods, game theories, behavior science, behavioral economics and finance, network sciences, and agent-based modeling and simulation, etc. As illustrated in [54, 9, 18, 32, 10, 3], such theories and methods can be used for EcoFin designs, developments, and decision-support, e.g.,

- modeling, simulating, representing, characterizing, designing and evaluating EcoFin and FinTech mechanisms, markets, products, participants, rules and governance, etc.;
- simulating and determining their working mechanisms, system characteristics and complexities;
- characterizing psychological, emotional and behavioral factors and influence;
- resolving issues and conflicts; and
- validating and reinforcing rules, organizational structures, processes, interactions and performance; etc.

**Quantitative methods.** Quantitative methods including mathematical and statistical modeling are widely applied in EcoFin designs and businesses. Economics and finance heavily rely on typical quantitative methods, including various numerical methods, time-series analysis, signal analysis, statistical including Bayesian methods, random methods, utility theories, and econometric methods. They are used in quantifying EcoFin businesses, e.g., formulating pricing and valuation models; quantifying the financial variables describing businesses; developing quality and performance evaluation matrices; evaluating and balancing supply and demand relationships; testing hypotheses and mechanisms; and analyzing issues such as uncertainty and chaos in the businesses; etc. In addition, more advanced analytics and learning methods such as deep learning methods are also used in quantitative finance and financial forecasting [51]. Interested readers can refer to [1, 9, 5, 14], etc. for more information on these methods and their applications in EcoFin and smart FinTech.

**Intelligent interactions, recognition and response.** EcoFin businesses are embedded into personal daily life and societal activities, where interactions (including dialogue and communications) between people, between humans and machines, and between machines become essential and overwhelming. Such interactions may take form of verbal dialogue (speech-based), textual communications (e.g., hand-written, emails, questions/answering, chatbots, or retrieval), visual dialogue (video-based), imagery interactions (e.g., images or QR codes), sensor-based (e.g., IoT devices-based) interactions, or multi-channel, multi-domain or multi-modal (e.g., audio-video plus chat) dialogue in EcoFin or by FinTech. Other businesses and tasks built on the above interactions include automated identification of interactions and communications and response generation for autonomous, open-domain, context-aware, multi-turn, multi-level, multi-lingual and knowledge-based dialogue, interactions and communications. Examples are [36,38,17,4]:

- automatically recognizing the identity of interacting objects from oral, image, video, retrieval or text-based dialogue, communications or queries;
- understanding interaction intent;
- analyzing sentiment and emotion;
- tracing historical dialogue context and feedback; and
- generating and recommending follow-up engagements and actions for automated responses, etc.

**Analytics and learning methods.** Both classic and advanced analytics and learning methods play increasingly critical roles in making EcoFin and FinTech smarter, more tailored, efficient, adaptive or evolving. Widely used classic methods consist of pattern analysis methods, kernel methods, tree models, factor models, relation models, representation learning, language models, image processing methods, signal pro-

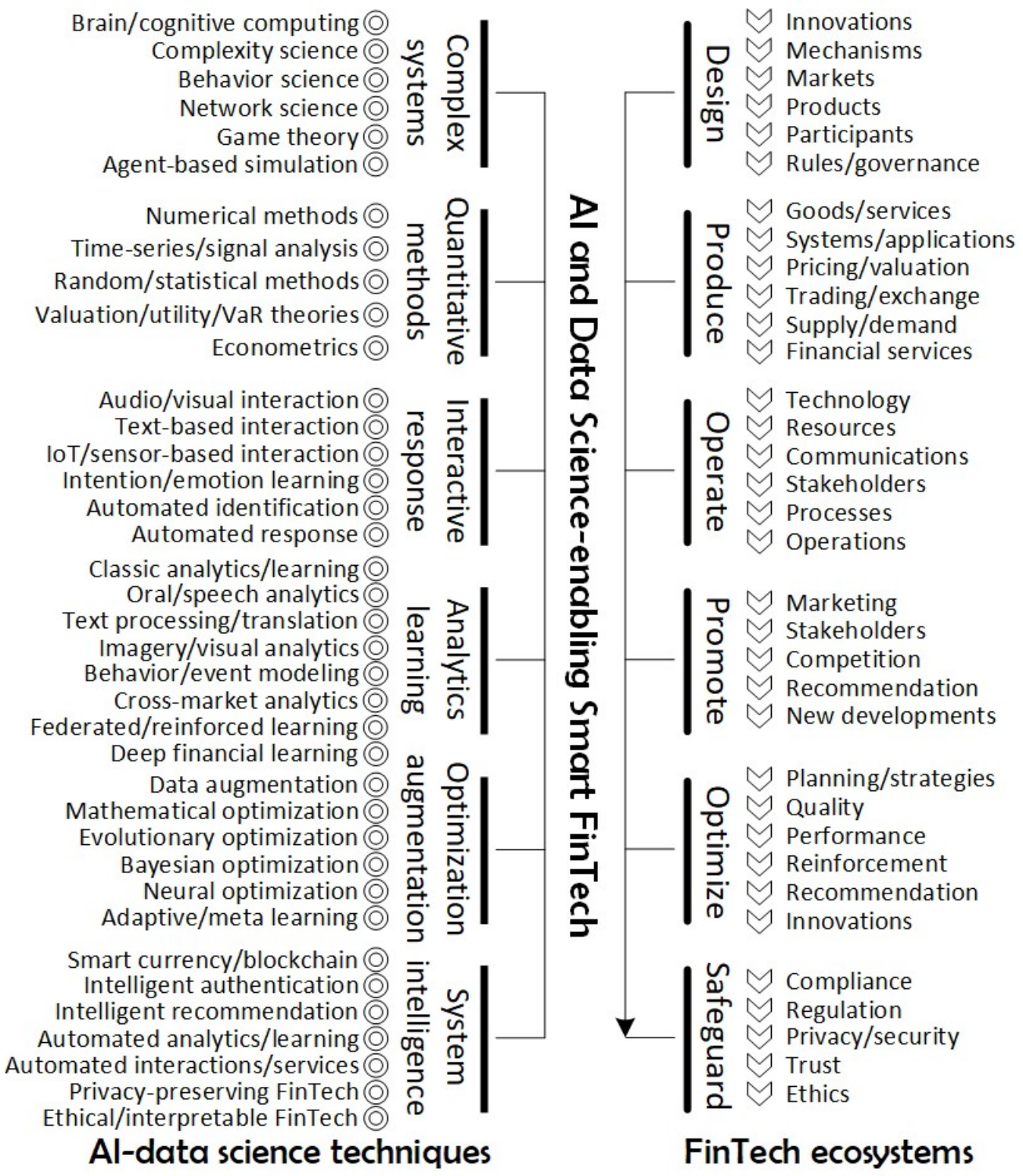

**Fig. 2** AI and data science enabling smart FinTech for financial businesses, processes and activities.

cessing techniques, classic neural networks, and ensemble methods, etc. More advanced analytics and learning dominates today's financial data analysis, financial engineering, and FinTech innovations. Such techniques comprise knowledge representation methods, short-long and informal-formal text processing and translation, imagery analytics, visual analytics, social media and network analysis, advanced optimization methods, advanced machine learning methods (e.g., on multimodal, multisource, high-dimensional, low-quality data, etc.), cross-business (e.g., products, policies, markets or customer groups) analytics, and whole-of-business (e.g., all products, services and customers of a bank) representation, analysis, recommendation and reinforced services. Numerous applications of these analytics and learning techniques have been reported in the literature and are possible, e.g., as discussed in [8, 9] and Table 1. Examples are

- predicting market trends or price movements such as jointly learning multi-source data and heterogeneous information fused together with financial events or social media sentiments [60];
- generating algorithmic trading strategies such as analyzing relevant price-sensitive financial events from the news and announcements about the security or predicting security price movement [41];
- detecting illegal or incompliant behaviors and events such as discovering pool manipulation from collective investment accounts and investor behaviors [8];
- modeling the impact of external events in financial markets such as the impact of information security on stock market movement [53]; and
- modeling cross-market investment strategies, investment herding behaviors, volatility spillover, and risk management [58,59].

**Deep financial modeling.** Deep financial modeling focuses on key issues that can be categorized into (1) deep financial representation and prediction, (2) deep cross-market, section and factor modeling, and (3) deep distributed financial modeling [9]. These tasks apply deep learning models, including basic deep neural networks and their recent developments, such as diversified deep neural mechanisms, architectures and networks (e.g., recurrent neural networks, graph neural networks, neural language models like Transformer and BERT variants, image nets, attention networks, memory networks, adversarial learning, and autoencoders, etc.). In addition, new developments on deep reinforcement learning, deep Bayesian learning, deep transfer learning and deep federated learning have also intensively transformed the landscape of DSAI-driven finance and FinTech [46,57,50,52]. Examples are

- recurrent neural networks for multivariate time-series forecasting, temporal and sequential modeling of financial variables, and price or index trend forecasting [51,26,52];
- deep reinforcement learning such as deep Q-networks for stock forecasting, optimal portfolio management, high-frequency trading strategy development for algorithmic trading [37,34,30];
- deep neural networks such as recurrent, regression and graph neural networks and deep transfer learning models for representing macro to micro-level financial data or cross-sectional stock representation for portfolio prediction and strategy development [43,46,55];
- deep learning of illegal, noncompliant, risky and fraudulent behaviors such as insider trading and market manipulation in capital markets and financial accounting and reporting fraud in financial services and statements [12]; and
- federated learning models for privacy-preserving open-domain and whole-of-business financial applications and services [19,57].

**Augmentation and optimization methods.** Augmentation and optimization techniques [29] may be useful to tackle business and data complexities in EcoFin businesses; enhance FinTech design objectives, mechanisms, capacity and strategies; or improve FinTech capability, quality and performance. EcoFin and FinTech may involve image, visual, audio, textual and transactional data with various quality issues such inconsistencies, noises, heterogeneity and feature hiddenness. Data augmentation techniques such as adding random noises, adversarial training, masking, generative adversarial networks and neural transformation may improve data quality and better serve FinTech objectives, e.g., contrast, translation, saturation and color augmentation for visual data; word shuffling, random swapping, insertion or deletion, and replacing synonyms for text; and cropping, changing speed and frequency and wavelet and Fourier transform for audio and time-series data. In addition, augmentation may enhance technical capability or analytical results, e.g., by adding distributional noises, biases or drifts to data or objectives, which also improve FinTech capacity and generality. On the other hand, optimization is increasingly essential for building robust and actionable FinTech, e.g., optimizing pricing and valuation, investment value at risk, transaction costs, portfolio management, and cross-market investment and risk management [11, 21]. Typical methods include mathematical optimization methods; statistical optimization such as Bayesian variational inference; information theory-driven optimization; neural optimization methods such as neural computing, evolutionary computing and fuzzy set theories; active and adaptive learning and meta-learning methods; and optimization methods for deep neural networks.

**System intelligence enhancement.** The smartness of FinTech not only relies on the above techniques but also requires extra system intelligence and its enhancement for smart EcoFin businesses and services. Here, we briefly highlight the following techniques to enhance the system intelligence of FinTech: smart blockchain, intelligent (re-)identification and authentication, automated analytics and learning, intelligent recommendation, tailored interactions and services, and ethical and interpretable FinTech [45,23,25].

- Smart cryptos and blockchain: for secure, distributed and automated infrastructures and systems; encrypted, efficient, robust and distributed valuation, investment, exchange and portfolio management of digital currencies (bitcoins and other cryptos); distributed and risk-tolerant trading strategies and contracting; adaptive risk and event management and regulation; and novel blockchain businesses and services.
- Intelligent (re-)identification and authentication: of identity, access, biometrics (eg., face, fingerprint, voiceprint, speaker or sign language), liveness (face or fingerprint) or from optical character recognition (OCR), dialogue systems, automated speech recognition systems, text-to-speech translation systems.
- Automated analytics and learning: of automatically collected and processed EcoFin data and businesses, updating EcoFin services or suggesting engagement opportunities or actions (such as alerts, responses, and recommendations), evidenced by validated analytical results.
- Intelligent recommendation: of EcoFin services and businesses (such as stocks, portfolios, financial plans, and promotions) tailored for specific clients by considering their needs, habits and intention of consumption, evolving circumstances, and feedback, etc.
- Intelligent interactions and services: with customers by interacting with them through automated, context-aware and personalized visual, verbal, imagery, textual or multi-modal dialogues or communications and by considering their circumstances, interests, intentions, demand and evolving context, etc.
- Privacy-preserving microservice architectures: breaks down complex large-scale enterprise-wide infrastructures to local services-oriented privacy-preserving federated components and microservices for efficient computing and services.

– Ethical and explainable FinTech: offers EcoFin businesses and services that are privacy-preserved, accountable, transparent, unbiased and interpretable, e.g., in FinTech and business designs, products, pricing, quality, accessibility, market campaigns, recommendations, and risk management.

**Whole-of-business and privacy-preserving federated FinTech.** With the networked and globalized world, economic activities from design to technical development, manufacturing, sales and trading, supply chains, cross-border and multi-national e-commerce, customer services and financial services are increasingly connected and chained. Such EcoFin businesses often involve multiple, large-scale, distributed and privacy-sensitive products, services, platforms, applications, markets, organizations, application domains, customer communities, regions, and data sources. An example is a smart residential district or property chain, where intelligent buildings connect housing services, strata services, utility services, carparking services, community services, shopping, and healthcare, etc. Such multi-aspect businesses require smart home/city-oriented FinTech to cater for whole-of-business, distributed and privacy-preserving business objectives, intelligent infrastructures, functionalities, operations, services, and stakeholder management. The hybrid systems also involve multi-source, heterogeneous, distributed, individual-to-organizational, and privacy-sensitive customers and their private activities, behaviors, interactions and preferences, which are recorded by Internet services, CCTV systems, mobile apps, and WiFi systems, etc. Whole-of-business and privacy-preserving federated systems [19, 57] and FinTech are highly demanding with functions such as securely linking all businesses; systematically serving the whole community; providing unified secure access, identification, authentication; supporting privacy-preserving and personalized interactions, conversations and responses within the community and between users; and providing multi-product/policy and cross-business offers and personalized recommendations. In a more general context, such whole-of-business and privacy-preserving federated FinTech and intelligent EcoFin businesses is able to

– connect all users, systems, organizations and data through heterogeneous, private-preserving and secure networks;
– fuse multi/cross-aspect and privacy-preserved businesses and services;
– support transparent (e.g., from heterogeneous sign-on apps) and privacy-preserving access, identification and authentication;
– offer privacy-preserved, connected, federated and differentiated financial systems and services;
– conduct online, cloud-based, privacy-preserving federated analytics and learning of all businesses, users and data;
– support customized, context-aware, multi-channel and instant interactions, communications and customer services;
– provide secure and personalized advertising, market campaigns, and recommendations.

Federated learning offers a decentralized method to train a machine learning model based on multi-party's data in a secure and privacy-preserving manner. It allows participating financial and non-financial organizations to collaborate without worry that their data privacy or key model parameters are leaked out unexpectedly. Often organizations' data and models complement each other either by enhancing the diversity of training samples or feature spaces. These dimensions allow machine learning models to be more robust and more accurate in financial applications. Many federated applications have been successfully implemented, in financial credit risk modeling, Internet marketing, call center automation, insurance and asset management [57,56].

## 5 About the edition on DSAI in FinTech

There are several initiatives on AI and data science in finance and FinTech, such as the special track on AI in FinTech organized in IJCAI'2020[2] and the panel on AI in FinTech: Challenges and future[3]. The special edition on Data Science and AI in FinTech is a major initiative of JDSA in promoting the research and applications of data science and AI in finance, the economy, society, and daily life for individual, public, economic and social good. The dozens of papers available online cover a wide spectrum of DSAI techniques and business domains and applications. Fig. 3 provides an initial picture of this collection of research as indicated by the frequently occurring keywords identified in these papers, particularly backed by the DSAA'2020 Journal Track on Data Science and AI in FinTech[4] and the special issue on the same topic[5]. Below, we briefly outline the scope and capacity of the early-stage collection of papers on this important topic.

**Research areas.** This special edition already covers a wide spectrum of research topics, tasks and methods. We summarize them in terms of the following top-ics: time series transformation, data manipulation and augmentation, financial forecasting and prediction, pattern mining, clustering and classification, behavior and event analysis, causality analysis, relation learning, network analysis, anomaly and fraud detection, emotion and sentiment analysis, influence and impact modeling, active, dynamic, online and adaptive learning, deep learning, performance evaluation, and other topics.

[2] https://static.ijcai.org/2020-accepted_papers.html
[3] https://ijcai20.org/ai-in-fintech-challenges-and-future
[4] http://dsaa2020.dsaa.co/journal-track-papers/
[5] https://www.springer.com/journal/41060/updates/19117592

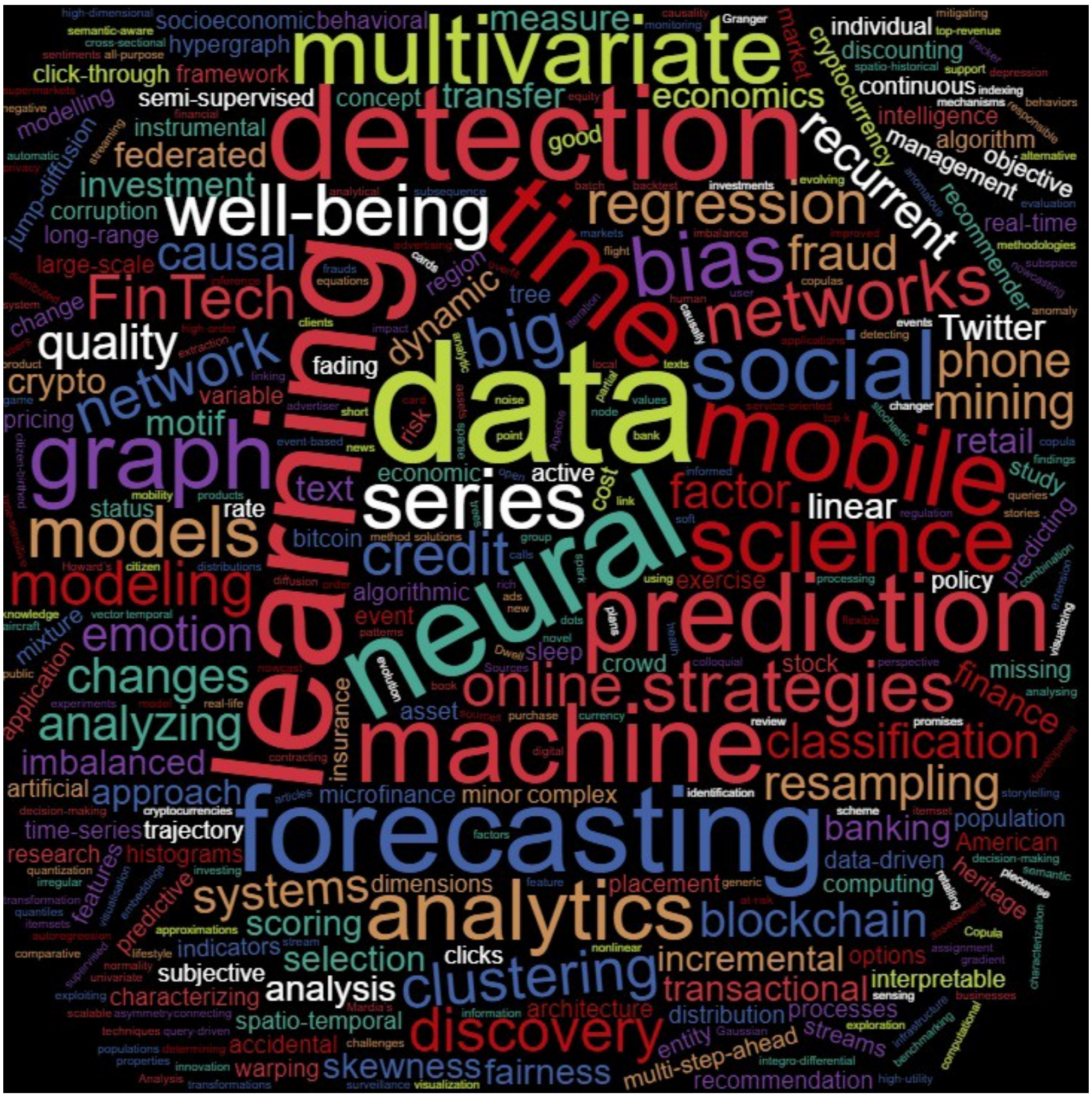

**Fig. 3** The word cloud of the collection on DSAI in FinTech.

– Multivariate time series transformation: including the transformation, modeling and representation of discrete and continuous multivariate time series (e.g., by copula transformations); handling multicollinearity; factor analysis; partial least squares regression; structural equation modeling; dynamic time warping and subsequence extraction.
– Data manipulation and augmentation: including handling data complexities such as high-dimensional, high-order, sparse, incomplete, hierarchical, cross-sectional, long-range, asymmetric, skewed, evolving, large-scale, imbalanced, live (real-time) and streaming data; handling noise and missing values; feature extraction and engineering; dimension reduction; sampling and resampling methods; visualization of spatial (altitude), temporal and historical data.
– Financial forecasting and prediction: including multivariate time-series analysis such as vector regression; cross-sectional short-to-long-range forecasting of multivariate financial variables by autoregression; classic classifiers, query-driven statistical learning, and resampling with temporal and relevance bias for time-series forecasting; biased resampling for imbalanced spatio-temporal forecasting; multi-step-ahead nonlinear multivariate forecasting; jump-diffusion models for high-order estimation of options prices; predicting socioeconomic status by hypergraph-based factor graph modeling of mobile call records.
– Pattern mining: such as high utility itemset mining of purchase records.

- Clustering and classification: of large-scale behavioral data (e.g., human actions and interactions) and blockchain bitcoin networks; cloud-based scalable clustering; diffusion-based region clustering; discovering time-series motifs (subsequences) by dynamic time warping; etc.
- Behavior and event analysis: including user modeling of online and social behaviors and people's lifestyle; extraction and classification of financial events from news articles; trajectory prediction by semantic clustering and neural network regressors; analyzing patterns of human behaviors and group interactions; complex event processing by detecting event and service patterns; consumer purchase behaviors.
- Causality analysis: such as linear and nonlinear Granger causality; causally anomalous multivariate time series; causal tree-based causal inference with instrumental variables; etc.
- Relation learning: including multivariate dependency (e.g., by copula methods), input-output causality (e.g., by Granger causality), query-answer relations, connectivity (e.g., distance, degree, centrality and densification, etc.) in bitcoin user-based graphs, and correlation skewness and multivariate skewness.
- Network analysis: including social media-based networks by social network and social media analyses; cryptocurrency and blockchain networks by graph theories; mobile phone calls-based networks by node embeddings.
- Anomaly and fraud detection: by causal anomaly detection, clustering methods, classification, incremental learning, transfer learning, network science methods, etc.
- Emotion and sentiment analysis: by social media analysis, topic modeling, text stream classification, visualization of sentiments, etc.
- Influence and impact modeling: such as the quantitative estimation of social and economic impact of data science on society, economy, industry and business; influence diffusion and propagation; socioeconomic development and well-being; measuring subjective and objective well-being; individual socioeconomic status; etc.
- Active, dynamic, online and adaptive learning: including online preprocessing and optimization of news data and entity prediction; exploratory active learning; stochastic semi-supervised learning; oversampling; multiple instance learning; incremental dynamic factor analysis.
- Deep learning: including interpretable neural networks to visualize financial texts; link prediction for credit scoring by recurrent neural networks, graph neural networks and autoencoders; sequential modeling and attention mechanisms of predicting data quality issues.
- Performance evaluation: including evaluating the overfitting, bias, usability, utility, representativeness, understandability, interpretability, explainability, transparency, accountability and liability of results.
- Other topics: including transfer learning and incremental learning of financial frauds; recommendations of individual and group purchase patterns by mixture models of point processes; online change detection in stochastic processes; distributional stream drifts and concept changes in business transactions; and imbalanced learning.

**Business domains and applications.** Many business problems, objectives, and applications have been addressed in this edition. They cover issues including (1) various business domains such as investment and trading, banking, retail industry and blockchain (digital currency, bitcoin, and cryptocurrency), pricing and valuation, and recommendation; (2) understanding the complexities of businesses, participants and their behaviors, such as population and community, customer behaviors and lifestyle, mobility, dynamics and change, financial events, information exchange, public sentiment and emotion, sectional and global businesses, socioeconomic impact, and cause-effect; and (3) supporting compliance and regulation and other issues, such as market surveillance, risk, data quality issues, fairness and ethics.

- Investment and trading: including equity investment, stock trading strategy backtesting, investment management, energy consumption, retail payment, sales of appliances, credit scoring.
- Banking: such as credit card fraud detection, credit scoring, and processing data quality issues.
- Retail businesses: including discovering high-utility products in supermarket transactions for revenue ranking and product indexing and placement, and region-specific product purchase patterns.
- Blockchain: such as modeling the relations between bitcoin owners and the concentration of user richness, and estimating cryptocurrency asset pricing.
- Community and citizen: such as minor heritage assets; population indications (residents, commuters and visitors) in mobile phone calls; public well-being and social good related to health, job opportunities, socioeconomic status, socioeconomic development, safety, environment and politics.
- Customer behaviors and lifestyle: such as individual and group purchase behaviors and preferences; user

resting, sleep, wakeup and exercise behaviors and quality; online and social behavior patterns; and accidental clicks of mobile ads and online advertising click-through.
- Mobility: such as on aircraft flight plans and trajectories, and mobile phone call-based human mobility.
- Dynamics and change: such as user behavior changes, and live stream drifts in multivariate time series.
- Financial events: such as social and public events, and web service events and patterns; and storytelling by generating stories by extracting and predicting entities and events in financial news articles.
- Information exchange: such as citizens as information providers and sentiment flow in social media.
- Public emotion: such as social sentiment, negative sentiments (e.g., depression, lack of interest, guilt, suicide, etc.) of at-risk populations, and market mood; and in online communities, and public communities.
- Cause-effect: such as evaluating heterogeneous causal effects (e.g., credit for small firms) by causal inference.
- Market surveillance: such as monitoring order behaviors in stock market order book and multivariate time series by causal outlier detection.
- Risk: such as fraudulent behaviors in banking etc. businesses, and corruption of contracting markets.
- Data quality issues: including identifying and processing diversified quality issues as discussed in the above on data manipulation and augmentation in financial businesses such as for banking regulation, and improving fairness and decision-support.
- Ethics and fairness: including privacy issues, selection bias in retail businesses, algorithmic bias in recommendations, temporal and relevance bias (imbalance), and fairness impact of biased missing values and selection bias.

Obviously, the above summary of the research areas and business applications in this edition will be substantially expanded over time.

## 6 Discussion and opportunities

Here, we discuss the trend of the era of smart FinTech and some opportunities for smart FinTech futures by the DSAI communities.

**The era of smart FinTech.** Smart FinTech is increasingly dominating and evolving into a paramount pillar of modern and future economies, society, and developments. FinTech connects every person, organization, product, service and activity anywhere, at any time and in any form through channels such as QR codes, WiFi networks, mobile applications, social media networks, short messaging platforms, intelligent digital assistants, and the Internet. They accumulate fast-growing tangible and intangible assets and services, numerous products, applications and services, and increasingly bigger financial data. AI and data science are playing an increasingly critical role in making finance smarter and fostering the ever-growing smartness of FinTech and the intelligence of autonomous financial systems and personalized financial services. To this end, DSAI advances include intelligent identification and authentication, autonomous interactions and communications, cloud analytics, deep learning, federated learning, cross-market analytics, deep financial modeling, and automated interactions and responses are also evolving to address the emerging challenges and opportunities in smarter financial businesses and bigger financial data.

**Opportunities to DSAI communities.** Finance is one of the mostly vivacious, data-rich, technology-savvy and intelligent areas. The advances of DSAI will further drive smart futures [9] of finance and FinTech, in areas including but not limited to:

- offering intelligent (e.g., proactive, optimal, autonomous, evidence-based, and differentiated) review, planning, generation and recommendations of financial objectives, areas, strategies, and actions;
- enabling anytime, anywhere and any-form access, identification, authentication, consumption, payment, communications, and responses to multiple heterogeneous networks and applications;
- enabling lifelong (evolving), all-purpose and whole-of-business (multi-policy) businesses and services to individual and group customers;
- suggesting novel opportunities such as new products, new functions, new markets, new customers, and new developments;
- providing whole-of-business, federated and privacy-preserving business support, data-driven learning, evidence extraction, recommendations and services;
- supporting global, all-purpose, cross-lingual, cross-modal, cross-business, and cross-apps privacy-preserving, transparent, tailored and active business fusion infrastructures, financial services, business recommendations and services;
- understanding user intention, emotion, preference, feedback and their evolution to offer positive, active and tailored customer experience and services; and
- automating integrative whole-of-process businesses and services, including autonomous product categorization, catalogue refilling, market positioning and promotion, billing, logistics dispatching, payment, product evaluation, and customer services, etc.

In addition, there are still many questions for the DSAI communities to answer, e.g.,

- how can AI and data science translate the existing finance and FinTech?
- what are the new smart futures of DSAI-driven finance and FinTech?
- what financial business and data complexities are not yet addressed by the existing analytics and learning family?
- how can DSAI algorithms autonomously access and analyze the whole-of-business data, generate whole-of-enterprise (or an individual object) data genomics, and make actionable recommendations for the management?
- how can DSAI enable efficient, privacy-preserving personal finance and Internet-based open banking and finance in a global or large-scale open domain?
- how can DSAI empower more transparent and robust digital currencies and blockchain systems and services?
- how has deep learning significantly transformed finance? and what are the essential finance and FinTech-oriented AI, deep learning and data science techniques?
- how can brain and cognitive computing and human-centered computing transform finance and FinTech?
- how can DSAI complement quantum computing for world-wide, secure and efficient financial infrastructures and services?
- to what extent can autonomous, personalized and precision finance possibly transform personal finance, corporate finance, public finance and Internet finance?

Table 1: DSAI Techniques and Their Representative Applications for Smart FinTech.

| DSAI areas | Techniques | Specific DSAI methods | Smart FinTech applications |
|---|---|---|---|
| Mathematical modeling | Numerical methods | Linear and nonlinear equations, least squares problem, finite difference methods, dependence modeling, Monte-Carlo simulation | Valuation, pricing, portfolio simulation, portfolio optimization, capital budgeting, hedging, price movement prediction, risk modeling, and trend forecasting. |
| | Time-series and signal analysis | State space modeling, time-series analysis, spectral analysis, long-memory time-series analysis, non-stationary analysis | Price prediction, trend forecasting, market movement prediction, IPO prediction, equity-derivative correlation analysis, change detection, financial crisis analysis, trading strategy discovery, and cross-market analysis. |
| | Statistical learning methods | Factor models, stochastic volatility models, copula methods, nonparametric methods, Bayesian networks | Market trend forecasting, pricing, valuation, price estimation, VaR forecasting, financial variable dependency modeling, portfolio performance estimate, and cross-market analysis. |
| | Random methods | Random sampling, random walk models, random forest, stochastic theory, fuzzy set theory, quantum mechanics | Abnormal behavior analysis, outlier detection, market event analysis, market movement modeling, influence transition analysis, associated account analysis, crowdsourcing modeling, and marketing modeling. |
| Complex system methods | Complexity science | Systems theory, complex adaptive systems, chaos theory, random fractal theory | System complexity modeling, market simulation, market mechanism design, globalization analysis, crisis contagion, and market information flow. |
| | Game theory | Zero-sum game, differential game, combinatorial game, evolutionary game, Bayesian game | Interaction modeling, policy simulation and optimization, regional conflict modeling, mechanism testing, coalition formation, and cryptocurrency mechanism testing. |
| | Network science | Linkage analysis, graph methods, power law, small worlds, contagion theory | Modeling entity movement, community formation, interactions and linkage, influence and contagion propagation; pool manipulation analysis, and analyzing investor relations. |
| | Agent-based modeling | Multiagent systems, belief-desire-intention model, reactive model, swarm intelligence, reinforcement learning | Testing economic hypotheses, simulating policies, supply chain relation modeling and optimization, cooperation analysis, self-organization modeling, portfolio optimization, and reinforcement learning. |
| Classic analytics and learning | Pattern mining methods | Frequent itemset mining, sequence analysis, combined pattern mining, high-utility pattern mining, tree pattern, network pattern, knot pattern, interactive pattern | Trading behavior analysis, abnormal trading analysis, outlier detection, investor relation analysis, customer profiling, high-utility trading pattern analysis, and cross-market trading behavior analysis. |
| | Kernel learning methods | Vector space kernel, tree kernel, support vector machine, spectral kernel, Fisher kernel, nonlinear kernel, multi-kernel methods | Price and market movement prediction, cross-market time series analysis, financial crisis analysis, crowdfunding estimate, market dependency modeling, and customer profiling and classification. |

Continued on next page

Table 1 – Continued from previous page

| DSAI areas | Techniques | Specific DSAI methods | Smart FinTech applications |
|---|---|---|---|
| | Event and behavior analysis | Sequence analysis, Markov chain process, high-impact behavior, high-utility behavior, nonoccurring behavior analysis | Financial event analysis, investor behavior analysis, price co-movement prediction, abnormal behavior analysis, market exception and change analysis, market event detection, and group behavior analysis. |
| | Document analysis and NLP | Language models, case-based reasoning, statistical language model, Bayesian model, latent Dirichlet allocation, Transformer, BERT | Financial event analysis, investor sentiment analysis, company valuation, financial review and auditing analysis, misinformation and rumor analysis, automated question-answering, and keyword searching. |
| | Model-based methods | Probabilistic graphical model, Bayesian networks, expectation-maximization model, clustering, classification, deep neural models | Hypothesis testing, customer clustering, price prediction, trend forecasting, index modeling, event analysis, fraud detection, movement forecasting, valuation, and risk scoring and prediction. |
| | Social media analysis | Topic modeling, sentiment analysis, emotional analysis, influence analysis, linkage analysis, interaction learning | Social influence analysis, investment linkage analysis, associated account analysis, sentiment analysis, influence modeling, market and price movement, detecting manipulation and insider trading, and understanding company branding and development. |
| Optimization and augmentation | Mathematical optimization | Nonlinear, stochastic and dynamic programming, information theory, Bayesian optimization | Optimizing policies, portfolios, trading strategies, VaR, and market performance, cost-benefit optimization. |
| | Neural computing methods | Wavelet neural network, genetic neural network, recurrent neural network, deep neural network | Macroeconomic and microeconomic factor correlation analysis, valuation and pricing modeling, relation analysis, sequence modeling, portfolio optimization, and trend and movement prediction. |
| | Evolutionary computing methods | Ant algorithm, genetic programming, self-organizing map, artificial immune system, swarm intelligence, neural-genetic algorithm | New product simulation, financial objective optimization, portfolio optimization, marketing strategy optimization, price and policy testing, market risk analysis, and market performance optimization. |
| | Fuzzy set methods | Fuzzy set theory, fuzzy logic, fuzzy neural network, genetic fuzzy logic | Modeling market momentum, financial solvency analysis, risk and capital cost modeling, and market uncertainty modeling. |
| | Data augmentation | Audio augmentation, image and visual augmentation, textual augmentation | Enhancing low-quality financial data, transforming data capacity, discovering latent features in data |
| Advanced analytics and learning | Representation learning | Probabilistic model, graph network, network embedding, tree model, neural embedding | Representation of stocks, assets, capital markets, portfolios, financial events, behaviors, and financial reports. |
| | Text processing and translation | Conceptualization, similarity learning of terms, tags and phrases, dependency parsing, word embedding, deep neural models, language translation models, neural translation networks | Text-based trend forecasting of price, market, sentiment and reputation, question/answering modeling and recommendation, bilingual or multilingual translation. |

Continued on next page

Table 1 – Continued from previous page

| DSAI areas | Techniques | Specific DSAI methods | Smart FinTech applications |
|---|---|---|---|
| | Reinforcement learning | Bellman Equation, actor-critic model, Markov dynamic progress, deep Q-network, adversarial reinforcement learning | Simulating and optimizing supply/demand of new assets and services, optimizing portfolios and trading strategies, option valuation optimization. |
| | Deep learning methods | Convolutional neural network, attention network, generative adversarial network, autoencoder, deep Bayesian network | Market modeling, behavior modeling, trading modeling, risk analysis, price and movement prediction, financial event modeling, cross-market analysis, cross-sectional modeling |
| | Federated and transfer learning | Multi-task learning, cross-domain transfer learning, distributed machine learning, federated learning, privacy-preserving processing, etc. | Offering privacy-preserving and user information-protected financial user interactions, dialogue, response, advertising, recommendation |
| Intelligent conversation and response | Intelligent identification and interactions | Imagery dialogue, visual dialogue, verbal/speech recognition, textual communications, multi-modal dialogue, IoT devices-based recognition, retrieval and question/answering, etc. | Automated recognition, dialogue and follow-up response with financial customers and stakeholders; recognizing their identities, liveness, biometric prints, voices, images, and visual actions for customer services and communications, etc. |
| | Intent and emotion learning | Attraction modeling, intent modeling, emotion learning, sentiment analysis, feedback analysis, etc. | Understanding customer intention, evolving demand and negative feedback on financial products and services, tailoring active interactions with providers, and recommending future services, etc. |
| System intelligence enhancement | Smart blockchain | Secure and efficient encryption, distributed ledger technology, smart contract protocols, user behavior analysis, anti-intrusion and fragility, etc. | Secure, distributed, efficient and risk-free infrastructure, cryptocurrencies and financial services (e.g., bitcoin portfolio management), etc. |
| | Intelligent authentication | Identity and object recognition, object segmentation, signal processing, visual analytics, and textual translation, personal (re-)identification, etc. | Offering and consuming financial businesses and services (e.g., online and short-messaging-based commerce) by mobile applications, in the IoT environment or WiFi networks, etc. |
| | Intelligent recommendation | Live recommendation, cold-start recommendation, tailed and sparse recommendation, personalized recommendation, cross-domain recommendation, next-basket/item recommendation, etc. | Customizing and suggesting the right products and services to the right people at the right time, etc. |
| | Automated learning | AutoML, active and adaptive learning, batch processing, automated data integration and processing, architecture search, context-aware learning, etc. | Automatically fusing, processing and analyzing heterogeneous financial information; automated performance optimization by optimal model search and updating, etc. |
| | Ethical FinTech | Feature coupling learning, representation disentanglement, explainable learning, visualization, unbiased learning, fairness learning, etc. | Offering unbiased, accountable and user-explainable products, prices, services, and recommendations, etc. |

Continued on next page

Table 1 – Continued from previous page

| DSAI areas | Techniques | Specific DSAI methods | Smart FinTech applications |
|---|---|---|---|
| Hybrid methods | Parallel ensemble | Evolutionary neural models, deep Bayesian model, copula graph neural network, combining complexity science and game theory | Price and market movement forecasting, multi-aspect risk analysis, macro/microeconomic factor analysis, financial event detection, customer profiling, blockchain and cryptocurrency modeling. |
| | Sequential and hierarchical hybridization | Time-series analysis plus classification, macro/microeconomic dependency modeling, deep sequential modeling-based event detection | Financial review-based fraud detection, macroeconomic influence on market movement, social media impact on price movement, epidemic evolution, and impact on index. |
| | Cross-disciplinary hybridization | Deep multi-time series analysis, copula deep models, autoregression deep model, behavioral economics and finance | Psychological factor and irrational market behavior analysis, behavioral finance, misinformation and mispricing on market inefficiency, financial ethics modeling, interpretable financial modeling. |
| | Behavioral economics and finance | Prospect theory, nudge theory, natural experiment, experimental economics, behavior informatics, intention learning, next-best action modeling | Modeling investment rationality, irrational behaviors, mispricing, market efficiency, mental activities, decision-making process, intention, emotion, and cognitive activities. |

## References

1. T. G. Andersen, R. A. Davis, J.-P. Kreiß, and T. V. Mikosch. *Handbook of financial time series*. Springer, 2009.
2. H. Arslanian and F. Fischer. *The Future of Finance: The Impact of FinTech, AI, and Crypto on Financial Services*. Palgrave Macmillan, 2019.
3. M. Baddeley. *Behavioural Economics and Finance*. Routledge, 2013.
4. A. Brahme and U. Bhadade. Effect of various visual speech units on language identification using visual speech recognition. *Int. J. Image Graph.*, 20(4):2050029:1–2050029:27, 2020.
5. L. Broemeling. *Bayesian Analysis of Time Series*. Chapman and Hall/CRC, 2019.
6. B. G. Buchanan. *Artificial intelligence in finance*. The Allen Turing Institute, 2019.
7. L. Cao. Data science: challenges and directions. *Commun. ACM*, 60(8):59–68, 2017.
8. L. Cao. AI in finance: A review. pages 1–35, 2020. https://ssrn.com/abstract=3647625.
9. L. Cao. AI in finance: Challenges, techniques and opportunities. pages 1–40, 2021. https://ssrn.com/abstract=3869625.
10. K. Chatterjee and W. F. S. (Eds.). *Game Theory and Business Applications*. Springer, 2001.
11. G. Cornuéjols, J. Peña, and R. Tütüncü. *Optimization Methods in Finance (2nd ed.)*. Cambridge University Press, 2018.
12. P. Craja, A. Kim, and S. Lessmann. Deep learning for detecting financial statement fraud. *Decis. Support Syst.*, 139:113421, 2020.
13. V. Dhar and R. M. Stein. Fintech platforms and strategy. *Commun. ACM*, 60(10):32–35, 2017.
14. C. Doloc. *Applications of Computational Intelligence in Data-Driven Trading*. John Wiley & Sons, 2020.
15. C. L. Dunis, P. W. Middleton, A. Karathanasopolous, and K. Theofilatos. *Artificial Intelligence in Financial Markets*. Springer, 2019.
16. N. Ehrentreich. *Agent-Based Modeling*. Springer, 2008.
17. M. Firdaus, N. Thakur, and A. Ekbal. Multidm-gcn: Aspect-guided response generation in multi-domain multi-modal dialogue system using graph convolution network. In *EMNLP'2020*, pages 2318–2328, 2020.
18. T. Fischer. News reaction in financial markets within a behavioral finance model with heterogeneous agents. *Algorithmic Finance*, 1(2):123–139, 2011.
19. B. C. M. Fung, K. Wang, R. Chen, and P. S. Yu. Privacy-preserving data publishing: A survey of recent developments. *ACM Comput. Surv.*, 42(4):14:1–14:53, 2010.
20. J. E. Gentle, W. K. Hardle, and Y. Mori. *Handbook of Computational Finance*. Springer, 2012.
21. M. Gilli, D. Maringer, and E. Schumann. *Numerical Methods and Optimization in Finance*. Academic Press, 2019.
22. I. Goodfellow, Y. Bengio, and A. Courville. *Deep Learning*. The MIT Press, 2016.
23. I. Hadi. Intelligent authentication for identity and access management: a review paper. *Iraqi Journal for Computers and Informatics*, 45(1):6–10, 2019.
24. L. Hamill and N. Gilbert. *Agent-Based Modelling in Economics*. John Wiley & Sons, 2016.
25. X. He, K. Zhao, and X. Chu. Automl: A survey of the state-of-the-art. *Knowledge-Based Systems*, 212:106622, 2021.
26. J. B. Heaton, N. G. Polson, and J. H. Witte. Deep learning for finance: Deep portfolios. *Applied Stochastic Models in Business and Industry*, 33:3–12, 2017.
27. Y. Hilpisch. *Artificial Intelligence in Finance*. OReilly, 2020.
28. International Telecommunication Union. Assessing the economic impact of artificial intelligence, 2018. Issue Paper No. 1, September 2018.
29. B. K. Iwana and S. Uchida. An empirical survey of data augmentation for time series classification with neural networks. *arXiv preprint arXiv:2007.15951*, 2020.
30. G. Jeong and H. Y. Kim. Improving financial trading decisions using deep q-learning: Predicting the number of shares, action strategies, and transfer learning. *Expert Syst. Appl.*, 117:125–138, 2019.
31. C. Kearney and S. Liu. Textual sentiment in finance: A survey of methods and models. *International Review of Financial Analysis*, 33:171–185, 2013.
32. K. Khashanah and T. Alsulaiman. Network theory and behavioral finance in a heterogeneous market environment. *Complexity*, 21(S2):530–554, 2016.
33. B. Kovalerchuk and E. E. Vityaev. *Data Mining in Finance: Advances in relational and hybrid methods*. Kluwer Academic Publishers, 2000.
34. K. Lei, B. Zhang, Y. Li, M. Yang, and Y. Shen. Time-driven feature-aware jointly deep reinforcement learning for financial signal representation and algorithmic trading. *Expert Syst. Appl.*, 140, 2020.
35. B. Li and S. C. H. Hoi. Online portfolio selection: A survey. *ACM Comput. Surv.*, 46(3):1–36, 2014.
36. H. Li, C. Huang, and L. Gu. Image pattern recognition in identification of financial bills risk management. *Neural Comput. Appl.*, 33(3):867–876, 2021.
37. Y. Li, P. Ni, and V. Chang. Application of deep reinforcement learning in stock trading strategies and stock forecasting. *Computing*, 102(6):1305–1322, 2020.
38. Z. Li, J. Kiseleva, and M. de Rijke. Improving response quality with backward reasoning in open-domain dialogue systems. In F. Diaz, C. Shah, T. Suel, P. Castells, R. Jones, and T. Sakai, editors, *SIGIR'2021*, pages 1940–1944, 2021.
39. T. Lynn, J. G. Mooney, P. Rosati, and M. Cummins. *Disrupting finance: Fintech and strategy in the 21st century*. Palgrave Pivot, 2019.
40. T. L. Meng and M. Khushi. Reinforcement learning in financial markets. *Data*, 4(3):110:1–17, 2019.
41. G. Mitra and L. Mitra. *The Handbook of News Analytics in Finance*. Wiley, 2012.
42. D. S. Modha, R. Ananthanarayanan, S. K. Esser, A. Ndirango, A. J. Sherbondy, and R. Singh. Cognitive computing. *Communications of the ACM*, 54(8):62–71, 2011.
43. K. Nakagawa, M. Abe, and J. Komiyama. RIC-NN: A robust transferable deep learning framework for cross-sectional investment strategy. In *DSAA'2020*, pages 370–379. IEEE, 2020.
44. B. Nicoletti and W. Nicoletti. *Future of FinTech*. Springer, 2017.
45. OECD. *OECD Blockchain Primer*. OECD, 2018. https://www.oecd.org/finance/OECD-Blockchain-Primer.pdf.
46. A. M. Özbayoglu, M. U. Gudelek, and O. B. Sezer. Deep learning for financial applications : A survey. *CoRR*, abs/2002.05786, 2020.
47. W. Pan and Q. Yang. Transfer learning in heterogeneous collaborative filtering domains. *Artif. Intell.*, 197:39–55, 2013.

48. Y. Qi and J. Xiao. Fintech: AI powers financial services to improve people's lives. *Commun. ACM*, 61(11):65–69, 2018.
49. L. Ryll, M. E. Barton, B. Z. Zhang, R. J. McWaters, E. Schizas, R. Hao, K. Bear, M. Preziuso, E. Seger, R. Wardrop, P. R. Rau, P. Debata, P. Rowan, N. Adams, M. Gray, and N. Yerolemou. Transforming paradigms: A global ai in financial services survey, 2020.
50. M. Sewak. *Deep Reinforcement Learning - Frontiers of Artificial Intelligence*. Springer, 2019.
51. O. B. Sezer, M. U. Gudelek, and A. M. Özbayoglu. Financial time series forecasting with deep learning : A systematic literature review: 2005-2019. *CoRR*, abs/1911.13288, 2019.
52. J. Sirignano and R. Cont. Universal features of price formation in financial markets: perspectives from deep learning. *Quantitative Finance*, 19:1449–1459, 2019.
53. G. Spanos and L. Angelis. The impact of information security events to the stock market: A systematic literature review. *Comput. Secur.*, 58:216–229, 2016.
54. M. M. Waldrop. *Complexity: The Emerging Science at the Edge of Order and Chaos*. Simon & Schuster Paperbacks, 1992.
55. Y. Wei and V. Chaudhary. The directionality function defect of performance evaluation method in regression neural network for stock price prediction. In *DSAA'2020*, pages 769–770, 2020.
56. Q. Yang, Y. Liu, T. Chen, and Y. Tong. Federated machine learning: Concept and applications. *ACM Trans. Intell. Syst. Technol.*, 10(2):12:1–12:19, 2019.
57. Q. Yang, Y. Liu, Y. Cheng, Y. Kang, T. Chen, and H. Yu. *Federated Learning*. Synthesis Lectures on Artificial Intelligence and Machine Learning. Morgan & Claypool Publishers, 2019.
58. J. Zhang and Q. He. Dynamic cross-market volatility spillover based on MSV model: Evidence from bitcoin, gold, crude oil, and stock markets. *Complex.*, 2021:9912418:1–9912418:8, 2021.
59. J. Zhang and Y. Zhuang. Cross-market infection research on stock herding behavior based on DGC-MSV models and bayesian network. *Complex.*, 2021:6645151:1–6645151:8, 2021.
60. X. Zhang, Y. Li, S. Wang, B. Fang, and P. S. Yu. Enhancing stock market prediction with extended coupled hidden markov model over multi-sourced data. *Knowl. Inf. Syst.*, 61(2):1071–1090, 2019.